\documentstyle[11pt,paspconf,epsf]{article}
\begin{document}

\title{Observational evidence for the accretion of low-metallicity gas onto the
Milky Way: \\metallicity, physical conditions and distance limit for HVC
complex~C}

\author{B.P. Wakker, J.C. Howk, B.D. Savage, S.L. Tufte, R.J. Reynolds}
\affil{Department of Astronomy, University of Wisconsin, Madison, WI 53706}
\author{H. van Woerden, U.J. Schwarz}
\affil{Kapteyn Institute, Rijks Universiteit Groningen, The Netherlands}
\author{R.F. Peletier}
\affil{University of Durham, UK}


\def\dex#1{10$^{#1}$}
\def\tdex#1{$\times$10$^{#1}$}
\def\Ha{H$\alpha$}
\def\HI{H{\small I}}
\def\HII{H{\small II}}
\def\SI{S{\small I}}
\def\SII{S{\small II}}
\def\CaII{Ca{\small II}}
\def\NHI{{\rm N(HI)}}
\def\NHp{{\rm N(H}\ifmmode^+\else$^+$\fi)}
\def\NSII{{\rm N(SII)}}
\def\cF{{\cal F}}

\begin{abstract}
We present observations of the (field of the) Seyfert galaxy Mark\,290, which
probes the high-velocity cloud (HVC) complex~C, one of the largest HVCs (Wakker
\& van Woerden 1991). We find that this object has a metallicity of
0.094$\pm$0.020$^{+0.022}_{-0.019}$ times solar. We determine a lower limit to
its distance of 5\,kpc ($z$$>$3.5\,kpc). If the gas is in thermal equilibrium
with a hot halo, then it is also likely that $D$$<$30\,kpc, putting the HVC in
the Galactic Halo, as was the case for HVC complex~A (for which $z$=2.5--7\,kpc;
van Woerden et al.\ 1998).
\par We find that, on this sightline, H$^+$ represents 23$\pm$10\% of the
hydrogen in the HVC. The total gaseous mass is 2.0\tdex6\ (D/5\,kpc)$^2$
M$_\odot$ and, depending on whether the space velocity is completely radial or
vertical, the HVC represents a downward mass flux of
$\sim$0.036--0.083\,(D/5\,kpc) \,M$_\odot$\,yr$^{-1}$, or
$\sim$2.9--6.7\tdex{-3}\,(D/5\,kpc)$^{-1}$ \,M$_\odot$\,yr$^{-1}$\,kpc$^{-2}$.
\par The low metallicity and large mass suggest that complex~C is unlikely to be
part of a Galactic Fountain, but rather represents accreting low-metallicity
material. It may be a present-day analogue of the damped Ly$\alpha$ absorbers
seen in QSO spectra. Our abundance result provides the first direct
observational evidence for the infall of low-metallicity gas on the Milky Way,
required in models of galactic chemical evolution.
\par It remains to be seen whether ultimately complex~C is a remnant of the
formation of the Milky Way (Oort 1970), a gas cloud orbiting the Galaxy (Kerr \&
Sullivan 1969), a Local Group object (Verschuur 1969, Blitz et al.\ 1996), the
result of tidal interactions between the Magellanic System and the Galaxy
$>$3\,Gyr ago (an "Old Magellanic Stream"; Toomre, quoted in Kerr \& Sullivan
1969), or was formed when hot, ionized intergalactic gas was compressed by the
motion of the Milky Way (Silk et al.\ 1987).
\end{abstract}
\eject

\section{Introduction}
In this contribution we describe results based on many observations of HVC
complex~C toward and near the Seyfert galaxy Markarian\,290 (V=14.96, z=0.030).
Figure~1 shows the large-scale \HI\ structure of the HVC this region. Among all
known extra-galactic probes of complex~C, Mark\,290 is the most favorable in
terms of the expected strength of absorption lines.
\par In Sect.~2 we discuss recent optical absorption-line data that allow us to
set a lower limit to the distance of the HVC. Sect.~3 describes the \SII\
absorption-line data, and in Sect.~4 we present data on optical and radio
emission lines. Sect.~5 describes a method for deriving an ionization-corrected
metallicity, ionization fraction, density and pressure for the cloud. This
method is applied in Sect.~6, and the implications are discussed in Sect.~7.
\par
\begin{figure}[t]
\plotfiddle{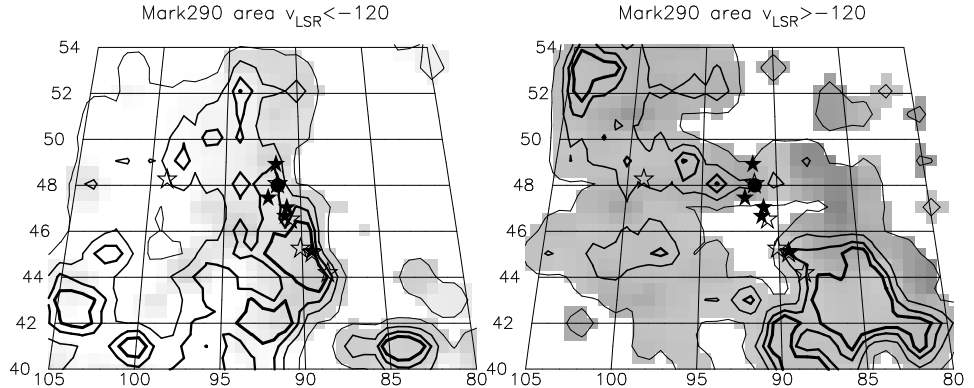}{5cm}{0}{80}{80}{-230}{-30}
\caption{\HI\ column density maps of part of complex~C (data from Hulsbosch \&
Wakker 1988). Brightness temperature contour levels are 0.05\,K and 25\%, 50\%,
75\% and 100\% of the peak value of the concentration nearest Mark 290. Filled
symbols indicate probe stars whose spectrum is included in Fig.~4. Larger
symbols are for more distant stars. Mark\,290 is shown by the filled circle.
}\end{figure}

\section{Distance limit}
Using the William Herschel Telescope (WHT) at La Palma, and the Utrecht Echelle
Spectrograph (UES), we observed \CaII\ H+K spectra of 12 stars in the region
around Mark\,290, at 6\,km/s resolution. The stars were selected from a list of
blue stars by Beers et al.\ (1996, and references therein). We chose stars that
were classified as Blue Horizontal Branch in low-resolution spectra; follow-up
spectroscopy confirms this for 80\% of such stars (Beers et al.\ 1992). For
these BHB candidates, a rough distance estimate can be made by assuming
$B$$-$$V$=0.05, and inserting this into the $B$$-$$V$ vs $M_V$ relation of
Preston et al.\ (1991). This gives $M_V$=0.86. Averaged over all possible
colors, the full range is $\pm$0.25 mag, so we calculated a probable distance
range using $M_V$=0.61 and 1.11. An extinction correction of $\sim$0.1 mag was
applied, using $A_V$ based on the map of Lucke (1978).
\par For the most distant stars the spectra are of relatively low quality, while
for some others stellar lines interfere with the detection of interstellar
features. Figure~2 shows five of the best \CaII\ K spectra, from which we can
set tentative limits on the strength of the Ca K absorption. The expected
equivalent widths can be derived from the \CaII/\HI\ ratio of 29$\pm$2\tdex{-9}
found toward Mark\,290 by Wakker et al.\ (1996), combined with the N(\HI) in the
direction of the star.
\par The estimated detection limit on the \CaII\ K absorption associated with
the HVC is always lower than the expected value. Since for three of the stars we
estimate D$\sim$5\,kpc, we conclude that the distance of complex~C is probably
$>$5\,kpc. However, there are three caveats.
\par a) The \HI\ column densities are based on a 9\arcmin\ beam, but
considerable variations at arcminute scales are possible (Wakker \& Schwarz
1991). Only a ratio (expected/detection-limit)$>$5 allows a safe conclusion.
\par b) The equivalent width limits are still preliminary.
\par c) The stellar distances need to be improved using spectroscopy.
\par
\begin{figure}[ht]
\plotfiddle{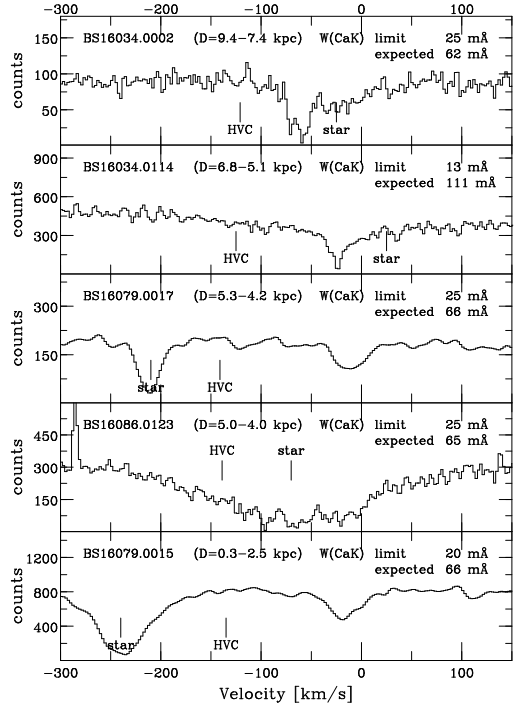}{9.5cm}{0}{80}{80}{-140}{-40}
\caption{
\CaII\ K spectra of five stars projected on complex~C. The range of possible
distances is shown, as are the detection limit and expected equivalent width.
The velocities of the HVC (from an Effelsberg spectrum) and the star (from
stellar features elsewhere in the spectrum) are indicated by the lines labeled
"HVC" and "star".
}\end{figure}

\section{Observations - \SII\ absorption}
\par We observed \SII\ absorption with the G140M grating of the Goddard High
Resolution Spectrograph (GHRS) on the Hubble Space Telescope (HST). Sulphur is
one of a few elements not depleted onto dust in the ISM, and S$^+$ is the
dominant ionization stage in neutral gas (Savage \& Sembach 1996). Thus,
N(S$^+$) allows a good measure of the intrinsic metallicity of an \HI\ cloud.
Among all similar ions, the \SII\ $\lambda\lambda$1250, 1253, 1259 lines are the
easiest to observe.
\par The integration time was 90 minutes, the resolution 15\,km/s. The \SII\
lines occur on top of strong Ly$\alpha$ emission associated with Mark\,290,
which increases the S/N ratio in the continuum at 1253\,\AA\ to 25, whereas it
is only 12 at 1259\,\AA. The $\lambda$1250 absorption is hidden by absorption
associated with Mark\,290.
\par The left top two panels of Fig.~3 show the spectra after continuum
normalization. The vertical lines correspond to zero velocity relative to the
LSR and to the two components observed at $-$138 and $-$115\,km/s in \HI\
(Wakker et al.\ 1996 and Fig.~3). The HVC component at $-$138\,km/s is clearly
seen in both \SII\ lines, but the $-$115\,km/s component is missing, although a
component at $-$110\,km/s may be present in the \CaII\ spectrum. This component
may be missing due to a combination of factors. First, it is wider (FWHM
31\,km/s vs 21\,km/s). Second, in the 9\arcmin\ beam it has a factor 1.6 lower
\HI\ column density. Third, fine structure in the emission may decrease the \HI\
column density even further; this is especially so for the $-$115\,km/s
component as Mark\,290 is near the edge of the $-$115\,km/s core (see Fig.~1).
In combination these may cause the \SII\ peak optical depth to become too low to
detect the line with the current S/N ratio.
\par Fitting the absorption lines between $-$165 and $-$125\,km/s gives
equivalent widths for the $\lambda$1253 and $\lambda$1259 lines of 20.3$\pm$3.4
and 32.7$\pm$5.2\,m\AA. Assuming no saturation and using $f$-values from Verner
et al.\ (1994), these correspond to column densities of 1.50$\pm$0.25\tdex{14}
and 1.63$\pm$0.30\tdex{14}\,cm$^{-2}$. The average, weighted by the S/N in the
continuum, is 1.54$\pm$0.27\tdex{14}. Half of this error is associated with the
placement of the continuum.
\par These absorption lines are unresolved, but likely to be unsaturated. We
base this conclusion on three lines of evidence. First, the \CaII\ lines toward
Mark\,290 were resolved by Wakker et al.\ (1996) (FWHM 14\,km/s at 6\,km/s
resolution), so the expected linewidth for \SII\ is 16\,km/s; an equivalent
width of 20.3$\pm$3.4\,m\AA\ would then give a column density of
1.6$\pm$0.3\tdex{14}\,cm$^{-2}$, in agreement with the value derived above.
Second, the observed ratio of equivalent widths is 1.6$\pm$0.4, which is
compatible with the expected ratio of 1.51. Third, for gas at temperature
$\sim$7000\,K (see Sect.~6), the thermal $b$-value for S is 1.9\,km/s; the
observed equivalent width then corresponds to an optical depth of 3.5 for the
$\lambda$1253 line. However, this predicts an equivalent width of 23.2\,m\AA\
for the $\lambda$1259 line, which is outside the error limit. If W(1259) were
1.51$\times$20.3$-$1$\sigma$=25.5\,m\AA, then for W(1253) to be 20.3\,m\AA\ one
requires $b$=2.9\,km/s with $\tau$=1.5 and 2.25 for the $\lambda$1253 and 1259
lines, respectively. This corresponds to a column density of
2.25\tdex{14}\,cm$^{-2}$, just 2.5$\sigma$ higher than the preferred value
above.
\par The \SII\ column density likely represents the total S column density. In
the ISM sulphur will exist as S$^+$ and either S$^0$ or S$^{+2}$. S$^0$ has an
ionization potential of 10.4\,eV, and thus is easily ionized by the ambient
radiation field. We do not detect \SI$\lambda$1262.86, setting a limit of
N(S$^0$)$<$2.6\tdex{14}\,cm$^{-2}$, or N(S$^+$)/N(S$^+$+S$^0$) $>$0.37. S$^+$
has an ionization potential of 23.3\,eV. If we assume that inside the part of
the cloud where H is fully ionized all S is S$^{+2}$ (which would imply that
there is no [\SII] emission), then, since we always find N(H$^+$)$<$N(\HI)
(Sect.~6), we conclude that N(S$^+$)/N(S$^+$+S$^{+2}$)$>$0.5. In low-velocity
neutral gas this ratio is always observed to be $>$0.9.
\par
\begin{figure}[ht]
\plotfiddle{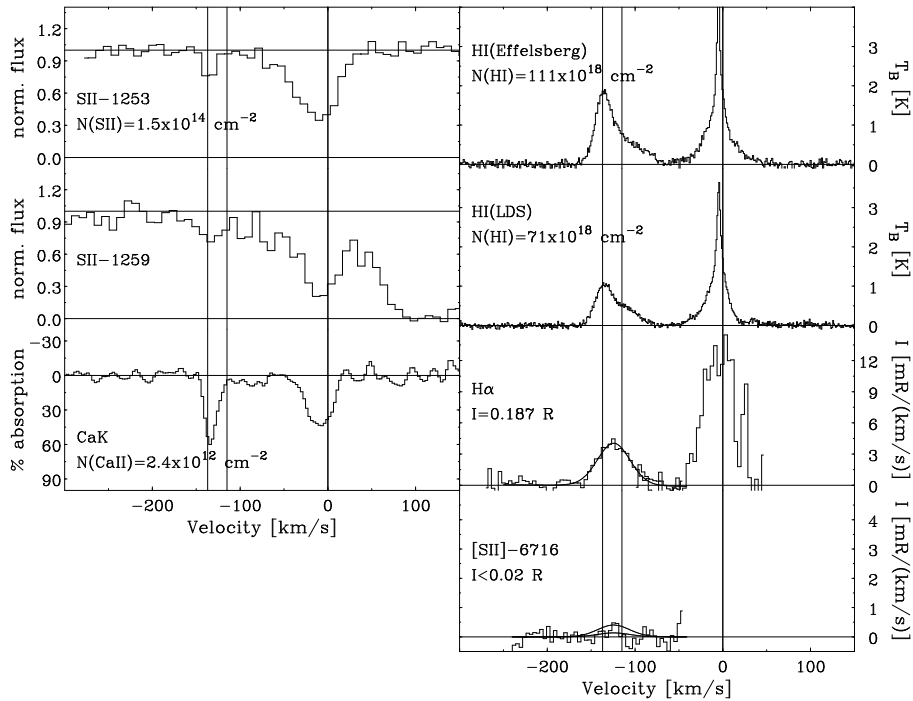}{9cm}{0}{75}{75}{-215}{-245}
\caption{
Spectra using Mark\,290 as a light bulb or centered on Mark\,290, aligned in
velocity. The labels give the spectral line and the resulting measured column
density or intensity for the HVC. For the [\SII] emission spectrum (WHAM data)
two curves are drawn, corresponding to 1/10th and 1/30th the \Ha\ intensity.
}\end{figure}

\section{Observations - emission lines}
\par To determine a metallicity from the \SII\ column density, \HI\ data with
the highest-possible resolution are required (see discussion in Wakker \& van
Woerden 1997). We have data from Westerbork at 1 arcmin resolution, but these
have not yet been analyzed. Until then we will use a 9 arcmin resolution
Effelsberg spectrum from Wakker et al.\ (1996). This spectrum is shown in
Fig.~3. There are two components, at $-$138 and $-$115\,km/s, with \NHI=68$\pm$3
and 43$\pm$7\tdex{18}\,cm$^{-2}$.
\par Using the Wisconsin \Ha\ Mapper (WHAM, Reynolds et al.\ 1998), we observed
the \Ha\ and [\SII] $\lambda$6716 emission in a one-degree diameter field
centered on Mark\,290. \Ha\ emission is clearly detected ($I_R$=0.187$\pm$0.010
R; where 1 Rayleigh is {$10^6/4\pi$} ph\,cm$^{-2}$\,s$^{-1}$\,sr$^{-1}$). The
spectrum in Fig.~3 has 12\,km/s resolution and is a combination of integrations
of 20 min for v$<$$-$50\,km/s and 30 sec for v$>$$-$100\,km/s. Most of the
``noise'' at lower velocities is fixed-pattern-noise.
\par [\SII] emission is not detected, although the 1-$\sigma$ noise level is
0.006\,R. Thus, we can set a 3-$\sigma$ limit of $<$0.1 for the ratio of \Ha\
and [\SII] intensities. WHAM usually sees a ratio of $\sim$0.3.
\par To compare the \Ha\ and \HI, an \HI\ spectrum of the full 1-degree WHAM
field is required, which was created from the Leiden-Dwingeloo survey (Hartmann
\& Burton 1997) by averaging the 4 spectra inside the WHAM beam. This spectrum
is also shown in Fig.~3 and gives column densities of 40.0$\pm$2.7 and
30.6$\pm$2.9\tdex{18}\,cm$^{-2}$ for the two \HI\ components. The apparent
velocity shift between \Ha\ and \HI\ is probably due to a different intensity
ratio between the two high-velocity components and the lower velocity resolution
of the \Ha\ data.

\section{Physical conditions - theory}
We now show how the \HI, \Ha, \SII\ absorption and [\SII] emission data can be
combined to derive an ionization-corrected S abundance ($A_{\rm S}$),
temperature ($T$), central density ($n_o$) and total ionization fraction ($X$).
We only need to assume a distance and a particular geometry, i.e.\ a density and
ionization structure.
\par We define the density structure in the sightline as $n(z)$ and the
ionization structure as $x(z)$, the ratio of ionized to total H (assuming
$n$(H$_2$)=0). The ``standard'' model consists of a cloud with diameter $L$ with
a fully neutral core with diameter $l$ and a fully ionized envelope with
constant density [$n(z)=n_o$ for $z$$<$$L/2$]. Figure~1 shows that the FWHM
angular diameter of the cloud, $\alpha$, is 2$\pm$0.2 degrees; the linear
diameter $L$ is the product of $\alpha$ and the distance ($D$).
\par To investigate the effects of different geometries we go one step beyond
this ``standard model'' and allow for a gaussian density distribution [$n(z)=
n_o\ \exp(-4\ln2\ z^2/L^2)$] and for the ionization fraction in the core and
envelope to be different from 0 or 1 [$x(z)$=$x_n$ for $z$$<$$l$; $x(z)$=$x_i$
for $z$$>$$l$]. We then need the following integrals: $$
\int_{-\infty}^\infty x(z)   n(z)  dz
            =  a         n_o  \ L\ [(1-r) x_n   + r x_i  ] =  \cF_1\ n_o  \ L
$$$$
\int_{-\infty}^\infty x(z)   n^2(z)dz
            = {a\over b} n_o^2\ L\ [(1-r) x_n   + r x_i  ] =  \cF_2\ n_o^2\ L
$$$$
\int_{-\infty}^\infty x^2(z) n^2(z)dz
            = {a\over b} n_o^2\ L\ [(1-r) x_n^2 + r x_i^2] =  \cF_3\ n_o^2\ L,
$$ where for a gaussian $r$=$1-{\rm erf}(\sqrt{4\ln2}\ l/L)$,
$a$=$\sqrt{\pi/4\ln2}$, $b$=$\sqrt{2}$, and for a uniform density distribution
$r$=$1-(2l/L)$, $a$=1, $b$=1. For the ``standard model'', all three $\cF$ values
reduce to the ``filling factor''.
\par The definition of \Ha\ emission measure and its relation to observables
are: $$
{\rm EM} = \int n_e(z)\ n({\rm H}^+)(z)\ dz
         = 2.75\ T_4^{0.924}\ I_R\ \ {\rm cm}^{-6}\,{\rm pc},
$$ with $T_4$ the temperature in units of 10000\,K, and $I_R$ the \Ha\ intensity
in Rayleigh.
\par For a cloud with substantial ionization, but still containing neutral gas,
most electrons will come from H, so that $n_e$=$\epsilon\,n({\rm H}^+)$, with
$\epsilon$$>$1. The first ionization potential of He is 24.6\,eV, so He will not
give a substantial contribution in mostly neutral gas. Because of their much
lower abundances all other elements combined contribute at most a few percent,
even if fully ionized. Thus, we have the following expressions for the H$^+$ and
\HI\ column density and the \Ha\ intensity in terms of the structure parameters:
$$
\NHp = \int_{-\infty}^\infty            x(z) n(z)   dz = \cF_1\ n_o\ L
$$$$
\NHI = \int_{-\infty}^\infty         (1-x(z))n(z)   dz = (a-\cF_1)\ n_o\ L
$$$$
2.75\ T_4^{0.924}\ I_R = {\rm EM} =
    \int_{-\infty}^\infty \epsilon\,x^2(z) n^2(z) dz = \epsilon \cF_3\ n_o^2\ L.
$$
\par To convert the observable (intensity) into the emission measure, we need to
know the gas temperature. This can be found by combining the S$^+$ emission and
absorption data. The ratio of S$^+$ and H$^+$ emissivity is:
$$
{{\rm \epsilon({\rm SII})}\over{\rm \epsilon({\rm HII})}}
      = 7.73\times10^5\ T_4^{0.424}\ \exp\left({-2.14\over T_4}\right)\
            \ \left({n_{S^+} \over n_{H^+}} \right)
      = F(T)\ \left({n_{S^+} \over n_{H^+}} \right).
$$ The density of S$^+$ is $n({\rm S}^+)(z) = A_{\rm S} n(z)$, with
$A_{\rm S}$ the S$^+$ abundance. Thus, locally, the emissivity ratio is some
constant times the density ratio. If we assume that S$^+$ emission occurs only
in the part of the cloud where electrons are present, and if we assume a
constant temperature, then the intensity ratio is:
$$
E         = {{\rm I({\rm SII})}\over{\rm I({\rm HII})}}
          =     F(T)\ {\int  n_e n_S    dz \over \int   n_e n({\rm H}^+)  dz }
          = A_{\rm S} F(T)\ {\int x(z) n^2(z) dz \over \int x^2(z) n^2(z) dz }
          = {\cF_2\over\cF_3}\ A_{\rm S} F(T).
$$ Our GHRS absorption measurement gives N(S$^+$) in the pencil beam to Mark
290. However, the [\SII] emission measure is determined by the column density
within the WHAM beam, which we estimate by scaling with the ratio of average
\NHI\ in the WHAM beam to \NHI\ in the pencil beam to Mark 290. So: $$
A_{\rm S} = {y\ \NSII \over \NHp + \NHI_{\rm WHAM}},\ {\rm with\ }
             y = {\NHI_{\rm WHAM} \over \NHI_{\rm Mark290}}.
$$
\par We now have five equations for the seven unknowns $T$, $\NHp$, $A_{\rm S}$,
$n_o$, $x_n$, $x_i$ and $r$, in terms of the observables \NHI$_{\rm WHAM}$,
\NHI$_{\rm Mark290}$, $\alpha$, $I$(\HII), $I$([\SII]), \NSII, and the distance.
We can solve this system using the following procedure. First, assume a
distance. Pick a value for $T$ to calculate EM(\HII). Assume $x_n$ and $x_i$,
and solve for $r$, $n_o$ and \NHp. From this calculate $A_{\rm S}$ and $E$.
Iterate until the derived and observed values of $E$ agree. Using the derived
values of $n_o$ and $T$, we can also calculate the pressure in the cloud as the
product $n_o T$.
\par
\begin{figure}[ht]
\plotfiddle{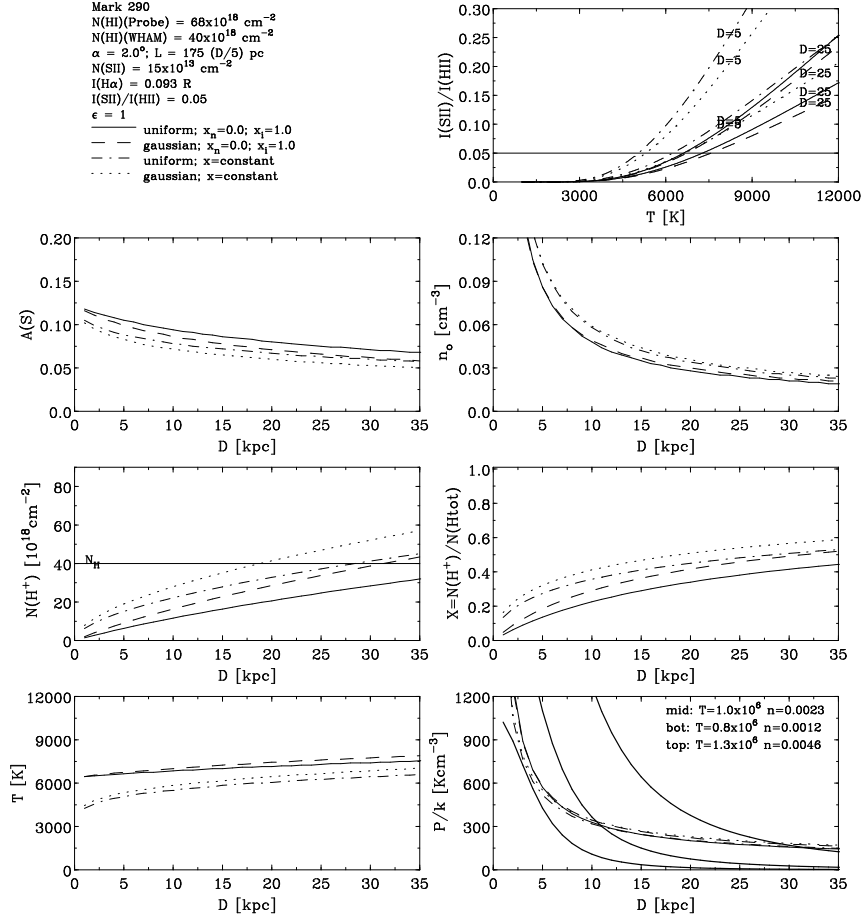}{11.5cm}{0}{64}{64}{-195}{-140}
\caption{
Derived values for gas temperature ($T$), S$^+$ abundance (A(S)), H$^+$ column
density (N(H$^+$)), ionization fraction ($X$), central density ($n_o$) and
central pressure ($P$), as a function of the unknown distance for four different
geometries (as indicated in the upper left panel).
}\end{figure}

\section{Physical conditions and metallicity - results}
We now derive the metallicity and physical conditions inside complex~C. We
calculate a reference value using the following assumptions: a) the geometry is
described by the ``standard model'' (constant density and a core-envelope
ionization structure); b) all electrons come from H ($\epsilon$=1); c) 50\% of
the \Ha\ emission is associated with the $-$138\,km/s component; d)
I([\SII])/I(\Ha)=1/20; e) there is no saturation in the \SII\ absorption; f) the
\SII\ absorption is only associated with the $-$138\,km/s \HI\ component; g) no
fine structure correction is needed; h) a distance of 10\,kpc.
\par With these assumptions, we can insert the observed values discussed in
Sects.~3 and 4 into the system of equations discussed in Sect.~5 to find a
sulphur abundance of 0.094$\pm$0.020 times solar. Here we used a solar abundance
of 1.86\tdex{-5} (Anders \& Grevesse 1989). With only the Effelsberg \HI\ data
and the \SII\ absorption column density we would have inferred an abundance of
0.121$\pm$0.022 times solar.
\par The error we give here is just the statistical error associated with the
measurements. It is dominated by the error in the \SII\ column density. We now
discuss the systematic errors introduced by the required assumptions.
%
%
%
\par A) Figure 4 shows the influence of four different assumptions for the
geometry: a uniform or a gaussian density distribution in combination with
either of two ionization structures: a neutral core and fully ionized envelope
($x_n$=0, $x_i$=1), or the same partial ionization throughout ($x_n$=$x_i$=$x$).
Changing the density structure to a gaussian results in an abundance of 0.086,
changing the ionization structure to constant partial ionization gives 0.078,
changing both gives 0.072. So, the possible variation in the metallicity
associated with geometry is $^{+0.000}_{-0.022}$. Of all uncertainties we
discuss this is the only one that cannot easily be improved upon with better
observations.
%
\par B) The assumption that all electrons come from hydrogen has little effect
on the results: if in the fully-ionized region He were also fully ionized
($\epsilon$$\sim$1.14), the derived abundance would be 0.097 ($+0.003$).
%
\par C) The \Ha\ emission is unresolved, so we cannot determine how much of the
emission is associated with each \HI\ component. If instead of 50\%, 25\% or
75\% of the \Ha\ emission is associated with the $-$138\,km/s component, the
abundance changes by $^{+0.007}_{-0.005}$. A higher angular resolution \Ha\
spectrum can reduce this uncertainty.
%
\par D) We did not actually detect [\SII] emission associated with complex~C,
but only have a (3$\sigma$) upper limit of $E$=I([\SII])/I(\Ha)$<$0.1. A deeper
observation of [\SII] is being planned. For the sake of the calculation we
assumed a value of 0.05. This uncertainty mostly influences the derived
temperature. For E in the range 0.01--0.10, the derived sulphur abundance varies
between 0.100 and 0.091, a range of $^{+0.006}_{-0.003}$.
%
\par E) If saturation is a problem for the \SII\ absorption lines, the column
density could be as high as 2.25\tdex{14}\,cm$^{-2}$, increasing the abundance
by 0.046.
%
\par F) A major problem is posed by the fact that the \HI\ spectrum shows two
components, while only one is seen in the \SII\ absorption spectrum. This
problem could be solved by a higher S/N and higher resolution observation of the
\SII\ absorption. If we use the total \HI\ column density of
111\tdex{18}\,cm$^{-2}$ (as well as the total \Ha\ emission), the derived
abundance is 0.061 times solar, a change of $-$0.033.
%
\par G) We do not yet know the precise value of \HI\ toward Mark\,290, as we 
used an Effelsberg spectrum with 9\arcmin\ beam. We will correct this to the
value for a 1\arcmin\ beam using Westerbork observations. By comparing to
similar cases (Wakker \& Schwarz 1991, Lu et al.\ 1998) we expect a correction
in the range 0.7--1.5. This changes the derived sulphur abundance to
0.137--0.062, a range of $^{+0.043}_{-0.032}$.
%
\par H) For an assumed range of distances from 5 to 25\,kpc, the change in the
derived metallicity is $^{+0.009}_{-0.019}$.
\par The ranges given above represent the largest deviations that can reasonably
be expected, equivalent to a 3-$\sigma$ errorbar. To calculate a combined
systematic error, we therefore add the ranges in quadrature and divide by 3. The
sources of uncertainty can be split into three groups: those associated with
physics (A-E) ($^{+0.016}_{-0.008}$), those associated with the \HI\ column
density (F,G) ($^{+0.014}_{-0.015}$), and the uncertainty associated with the
unknown distance (H) ($^{+0.003}_{-0.006}$). Combining these, the final value we
derive for the sulphur abundance in complex~C is
0.094$\pm$0.020$^{+0.022}_{-0.019}$ times solar.
\par Figure 4 shows the dependence on the assumed geometry and distance for the
derived parameters. Similar figures could be made showing the dependence on the
other assumptions. As was the case for the abundance, we can derive a fiducial
value and estimate the systematic error for the other parameters. We then find
that:
$X$=0.23$\pm$0.06$^{+0.07}_{-0.04}$, 
$n_o$=0.048$\pm$0.010$^{+0.011}_{-0.002}~\left(D/10\right)^{-0.5}$~cm$^{-3}$.
$T$=6800$\pm$500$^{+750}_{-900}$\,K, and
$P$=330$\pm$70$^{+105}_{-45}~\left(D/10\right)^{-0.5}$~K\,cm$^{-3}$.
%
%
%
%
%

\section{Discussion}

\subsection{Thermal pressure vs hot halo gas}
We can compare the derived thermal pressure with the thermal pressure of hot
halo gas. Wolfire et al.\ (1995) give a semi-empirical formula, using a base
density for the hot gas n(z=0)=0.0023\,cm$^{-3}$ and a temperature of order
\dex6\,K. From the ROSAT X-ray data, Snowden et al.\ (1998) find that the
probable value of the halo temperature is \dex{6.02\pm0.08}\,K. The emission
measure is about 0.02\,cm$^{-6}$\,pc (to within a factor of order 2), which for
the density given above corresponds to a scaleheight of order 5\,kpc. Such a
scaleheight would be similar to that observed for the highly-ionized atoms of
C$^{+3}$, N$^{+4}$ and O$^{+5}$ (Savage et al.\ 1997, Widmann et al.\ 1998).
\par The middle of the three bold-faced curves in the lower-right panel of
Fig.~4 shows the semi-empirical relation for n(z=0)=0.0023\,cm$^{-3}$, and
T=\dex{6.02}\,K. Both the pressure relation derived by Wolfire et al.\ (1995)
and the pressure derived by us represent the actual thermal pressure. Thus, if
there is pressure equilibrium, the most likely values of $n(z=0)$ and $T$ imply
a distance to complex~C of 10\,kpc. If the hot halo temperature were
\dex{5.94}\,K and the density were half as large, the implied distance is
$<$3\,kpc, incompatible with the observed lower limit. On the other hand, a
higher temperature (\dex{6.10}\,K) and density (double the value) would result
in equilibrium at a distance of 30\,kpc. We conclude that for reasonable values
for the density and temperature of hot halo gas, complex~C cannot be more
distant than $\sim$30\,kpc.

\subsection{Mass, energy and mass flow}
The mass of complex~C can be calculated by summing the observed column
densities, in the manner described by Wakker \& van Woerden (1991). We make the
assumptions that all the gas is at the same distance (unlikely, but we have
insufficient information to justify a different assumption), that N(H$_2$) can
be ignored (see Wakker et al.\ 1997), that N(H$^+$)/N(Htot)=0.23 everywhere, and
that there is a 28\% mass fraction of He. This yields a mass of
2.0\tdex6\,(D/5\,kpc)$^2$\,M$_\odot$.
\par To calculate the kinetic energy and mass flow associated with complex~C
requires some knowledge of its spatial velocity. Observed is the velocity
relative to the LSR, which contains the motion of the Sun and a contribution
from galactic rotation at the position of the object. To correct for these
contributions we use the deviation velocity (Wakker 1991), the difference
between the observed LSR velocity and the maximum LSR velocity that can be
easily understood in terms of differential galactic rotation. It is the minimum
velocity that the cloud has relative to its local environment. Integrating the
product of the mass and the square of the deviation velocity at each point in
the cloud and correcting for H$^+$ and He, leads to a kinetic energy of
$>$5.6\tdex{46}\,(D/5\,kpc)$^2$\,J, equivalent to the total energy of $>$500
supernovae.
\par We estimate the mass flow by making two different
assumptions: A) the space velocity is completely radial ($v_z=v_{\rm dev}\
\sin\,b$), or B) it is completely vertical ($v_z=v_{\rm dev}/sin\,b$). This
gives 0.036 and 0.083\,(D/5\,kpc)\,M$_\odot$\,yr$^{-1}$, respectively. The area
covered by complex~C is 1623 square degrees (12.4\,(D/5\,kpc)$^2$\,kpc$^2$), so
the corresponding infall rate is
2.9--6.7\tdex{-3}\,(D/5\,kpc)$^{-1}$\,M$_\odot$\,yr$^{-1}$\,kpc$^{-2}$. This
value is similar to the rate of $\sim$4\tdex{-3}\,
M$_\odot$\,kpc$^{-2}$\,yr$^{-1}$ required by models of galactic chemical
evolution (Giovagnoli \& Tosi 1995). However, the theoretical rate should be
present over the whole Galactic Disk, whereas the HVCs cover only $\sim$18\% of
the sky. More HVC distances and metallicities are needed to solve this possible
discrepancy.

\subsection{Origins}
\par Our metallicity excludes that complex~C is part of a Galactic Fountain, as
its metallicity then should have been $>$0.3 solar, the lowest value found in
the outer galaxy (Afflerbach et al.\ 1997). The HVC thus must be
extra-galactic.
\par Oort (1970) presented a model of continuing infall, in which gas originally
near the (inhomogeneous) transition region separating Milky Way gas from
intergalactic gas is still falling onto the Milky Way. This gas starts out hot
and ionized, and becomes visible after interacting, cooling and mixing with
high-z galactic gas associated with activity in the disk. Oort's model predicts
z-heights of $\sim$1\,kpc and metallicities of $\sim$0.7 times solar. The
distance limits for HVC complexes~A (z=2.5--7\,kpc) and C (z$>$3\,kpc) (van
Woerden et al.\ 1998 and in these proceedings) and our metallicity result for
complex~C imply that the clouds would have to become neutral at much higher z
than Oort's model suggests.
\par A Local Group origin for HVCs was first suggested by Verschuur (1969), who
noted that according to the virial theorem some of the then-known clouds would
be gravitationally stable at distances of 400\,kpc. However, for most of the
clouds found in later surveys the stability distance is several to tens of Mpc,
implying M$\sim$\dex{10}\,M$_\odot$. Thus, this idea was no longer taken
seriously. Blitz et al.\ (1998) point out that dark (and/or ionized) matter may
be present, so that \HI\ represents only 10--15\% of the total mass. This
reduces the average stability distance to 1\,Mpc. Based on this and many other
considerations, they suggest that the HVCs are remnants of the formation of the
Local Group. The large HVCs, including complex~C, would be nearby examples.
\par Extra-galactic HVCs could also represent gas orbiting the Milky Way, rather
than gas in the Local Group at large. This was originally suggested by Kerr \&
Sullivan (1969), who considered HVCs to be loosely bound intergalactic material,
too diffuse to contract into proto galaxies, orbiting the Galaxy at distances of
order 50\,kpc. They quote Toomre as suggesting that the source of this gas could
be tidal streams pulled out of the Magellanic Clouds during previous passages.
The metallicity would then be similar to that in the outer regions of the
Magellanic Clouds $>$5\,Gyr ago.
\par Mallouris et al.\ (1998) suggest that the HVCs are similar to Ly$\alpha$
absorbers. Vladilo (1998) suggests that damped Ly$\alpha$ absorbers are
associated with dwarf galaxies. Therefore, low-metallicity HVCs such as
complex~C could be failed dwarf galaxies, in which in the early universe some of
the gas formed stars, producing the observed metals, but where star formation
has currently stopped.

\end{document}